\documentclass[pre,twocolumn]{revtex4}
\usepackage{epsfig}
\usepackage{bm}
\usepackage{amssymb}
\usepackage{amsmath}

\begin{document}

\title{From deterministic to distributed chaos/turbulence in Rayleigh-B\'{e}nard convection: generalized Birkhoff-Saffman invariant }

\author{A. Bershadskii}

\affiliation{
ICAR, P.O. Box 31155, Jerusalem 91000, Israel
}

\begin{abstract}

 The transition from deterministic to distributed chaos/turbulence at the increase of Rayleigh number (from $10^4$ to $10^8$) in Rayleigh-B\'{e}nard convection, controlled by generalized Birkhoff-Saffman invariant, has been studied using the results of direct numerical simulations. The applications of this approach to rotating Rayleigh-B\'{e}nard convection, to stably stratified flows, and to observations in the atmosphere and in the solar photosphere have been briefly discussed.

\end{abstract}

\maketitle

\section{Inroduction}

 The Rayleigh-B{\'e}nard buoyancy-driven convection in a layer of fluid bounded below and above by two fixed temperature horizontal planes or a vertical temperature profile is one of the most interesting (both from theoretical and practical points of view) problems in fluid dynamics. In direct numerical simulations the fluid is usually taken as incompressible and the bounding horizontal planes as impenetrable walls with stress-free or no-slip boundary conditions for the velocity field, whereas the boundary conditions in the horizontal directions are taken as spatially periodic (that allows to mimic classic Rayleigh-B{\'e}nard convection typical for atmospheric and for solar convection). The existence of the side walls in closed containers, which is typical for laboratory experiments, can drastically change the physics of convection. This situation will be considered elsewhere. \\

In the paper Ref. \cite{hcl} three regimes of the Rayleigh-B{\'e}nard convection were introduced according to the experimental observations - chaotic (deterministic chaos), soft turbulence, and hard turbulence. The values of the Rayleigh number $Ra$ separating between these regimes depend on many other parameters and conditions (Prandtl number, aspect ratio, boundary conditions, etc). 

  The deterministic chaos was discovered in fluid dynamics by the means of a numerical simulation just for the  Rayleigh-B{\'e}nard convection \cite{lorenz}. The other two empiric regimes (soft and hard turbulence) were vaguely defined in the literature. There were attempts to relate the hard turbulence to the appearance of an inertial range of scales with power (scaling) spectral law. Indeed, deterministic chaos is usually associated with smooth trajectories and corresponding exponential power spectra \cite{fm}-\cite{kds}, whereas non-smooth dynamics is typically associated with the power-law (scaling) spectra. Smooth dynamics, however, can be generally characterized by the stretched exponential spectra. Therefore, there can be chaotic-like dynamics that is not deterministic chaos. This type of chaotic-like dynamics can be considered as a soft turbulence which separates the deterministic chaos and the hard turbulence. This type of dynamics is considered in the present paper using the notion of distributed chaos and generalized Birkhoff-Saffman invariant.

 \section{Deterministic chaos}

   In the paper Ref. \cite{ytc} results of the direct numerical simulations of the Rayleigh-B{\'e}nard convection in the plane layer (more precisely, in computational domains with a square cross-section and with large aspect ratios $\Gamma$) were reported for different values of the Rayleigh number $Ra$ (at the Prandtl number $Pr =1$). The system of equations (in the Boussinesq approximation) was taken in the form
 $$
\frac{\partial {\bf u}}{\partial t} + ({\bf u} \cdot \nabla) {\bf u}  =  -\frac{\nabla p}{\rho_0} + \alpha g T {\bf e}_z + \nu \nabla^2 {\bf u} ,  \eqno{(1)}
$$
$$
\frac{\partial T}{\partial t} + ({\bf u} \cdot \nabla) T  =   \kappa \nabla^2 T, \eqno{(2)}
$$
$$
\nabla \cdot \bf u =  0, \eqno{(3)}
$$   
 ${\bf u}$, $p$ and $T$ are the velocity, pressure and temperature fields, ${\bf e}_z$ is the vertical unit vector, $\rho_0$ is the mean density of the fluid, $g$ is the gravity acceleration, $\alpha$ is the thermal expansion coefficient, $\nu$ and  $\kappa$ are the viscosity and thermal diffusivity. \\ 
 
   This system was numerically solved with the impenetrable, stress-free, isothermal boundary conditions on the horizontal (plane) boundaries and periodic boundary conditions on the side boundaries (the aspect ratio is denoted as $\Gamma =L/H$, where $H$ is the distance between the two horizontal boundaries and $L$ is the periodicity length in both horizontal directions $x$ and $y$). The stress-free boundary conditions allow for avoiding the viscous boundary layer problems on the horizontal impenetrable boundaries.\\ 
 
   The main non-dimensional parameters characterizing the Rayleigh-B{\'e}nard convection are the Rayleigh number ${\rm Ra}=\sigma g \Delta T d^3/(\nu \kappa)$ and the Prandtl number $Pr=\nu/\kappa$ (where the temperature difference between the horizontal plane layer boundaries $\Delta T$ determines the energy input into the system).\\
 
   Figure 1 shows the horizontal kinetic energy spectrum computed at the midplane vs horizontal wavenumber $k_h = \sqrt {k_x^2 +k_y^2}$ obtained in the DNS for $Ra=10^4$, $Pr =1$ and $\Gamma =14.15$ (the spectral data were taken from Fig. 10d of the Ref. \cite{ytc}). The dashed curve indicates the exponential spectrum 
$$ 
E(k_h) \propto \exp(-k_h/k_c)  \eqno{(4)}
$$
  where $k_c$ is a characteristic value of the wavenumber, indicated by the dotted arrow in Fig. 1. \\
  
   It is known that the exponential spectrum is a typical characteristic of dynamical systems with smooth trajectories (deterministic chaos) \cite{fm}-\cite{kds}. Therefore, the kinetic energy spectrum shown in Fig. 1 can be considered a strong indication of deterministic chaos in this case as well. It should be noted that in the fluid dynamics the deterministic chaos was discovered just in the  Rayleigh-B{\'e}nard convection \cite{lorenz}.\\
   
   In the previous example, a case of the Rayleigh-B{\'e}nard convection at a small Rayleigh number was considered at the time of the simulation when it already reached a statistically stationary state. It is now interesting to consider an example of the Rayleigh-B{\'e}nard convection at a moderately large Rayleigh number $Ra = 10^7$ ($Pr = 0.71$, $\Gamma =2\pi$) but at an earlier time of DNS (at the onset of the convection) when it is still far away from the statistically stationary state \cite{par}. One can expect that at this stage of the convection development it is dominated by deterministic chaos (about further development of the convection see Section VI-B). \\
   
   Figure 2 shows the horizontal kinetic energy spectrum (vs horizontal wavenumber $k_h$) at an early time ($t =13$ in the DNS terms \cite{par}) of the convection development (the statistically stationary state for this DNS was reached at $t \sim 160$). It also should be noted that the no-slip boundary conditions were used for the DNS reported in the Ref. \cite{par}. The spectral data were taken from Fig. 3 of the Ref. \cite{par}. The dashed curve indicates the exponential spectrum (deterministic chaos). \\

\begin{figure} \vspace{-1.3cm}\centering
\epsfig{width=.45\textwidth,file=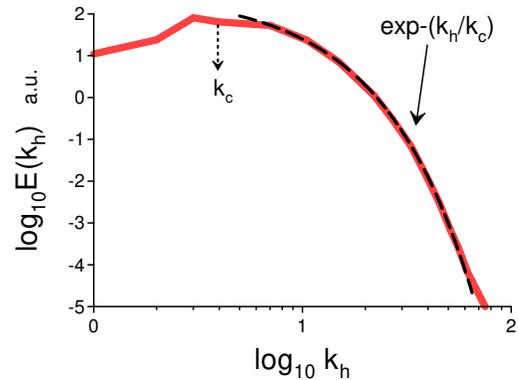} \vspace{-4.1cm}
\caption{Horizontal kinetic energy spectrum obtained in the DNS \cite{ytc} for $Ra=10^4$ at a statistically stationary state.} 
\end{figure}
\begin{figure} \vspace{-0.5cm}\centering
\epsfig{width=.45\textwidth,file=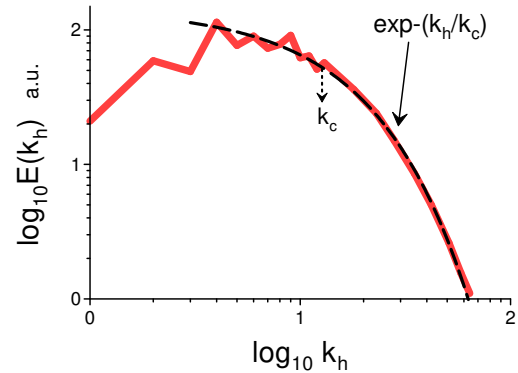} \vspace{-4cm}
\caption{Horizontal kinetic energy spectrum obtained in the DNS \cite{par} for $Ra=10^7$ at the onset of the DNS (far away from the statistically stationary state).} 
\end{figure}

\section{Generalized Birkhoff-Saffman invariant in Rayleigh-B\'{e}nard convection}
   
   While kinetic energy conservation takes place in ideal (non-dissipative) hydrodynamics it is broken by molecular viscosity (dissipation) for the Navier-Stokes equations. The Navier-Stokes equations have their own fundamental (dissipative) invariants related to the conservation of momentum and angular momentum - Birkhoff-Saffman and Loitsyanskii integrals \cite{my}-\cite{saf} (and due to the Noether theorem to homogeneity and isotropy, respectively).  \\
   
   The Birkhoff-Saffman invariant can be written as 
$$   
I_{BS} = \int  \langle {\bf u} ({\bf x},t) \cdot  {\bf u} ({\bf x} + {\bf r},t) \rangle d{\bf r}  \eqno{(5)}
$$ 
where $<...>$ denotes an ensemble (or volume) average (see, for instance, \cite{dav} and references therein).  \\

  The system of Eqs. (1-3) can be rewritten (by a change of variables) in the form
$$
\frac{\partial {\bf u}}{\partial t} + ({\bf u} \cdot \nabla) {\bf u}  =  -\frac{\nabla p'}{\rho_0} + \alpha g \theta {\bf e}_z + \nu \nabla^2 {\bf u}   \eqno{(6)}
$$
$$
\frac{\partial \theta}{\partial t} + ({\bf u} \cdot \nabla) \theta  =    \frac{\Delta T}{H} u_z + \kappa \nabla^2 \theta, \eqno{(7)}
$$
$$
\nabla \cdot \bf u =  0 \eqno{(8)}
$$
(see, for instance, Ref. \cite{kcv} and references therein). Here $p'$ and $\theta$  are the modified pressure and temperature fluctuation fields ($\theta = T-T_0 (z)$ where $T_0(z)$ is a linear vertical temperature profile), the temperature difference between the bottom and top boundaries is $\Delta T$. Usually, the periodic boundary conditions are applied on the side walls of the spatial domain, and the free-slip boundary conditions for velocity and $\theta =0$ on the bottom and top boundaries. \\   

  The Birkhoff-Saffman integral can be readily generalized on the buoyancy-driven motion Eqs. (6-8) as 
 $$
I_b =   \int  \langle {\bf u} ({\bf x},t) \cdot  {\bf u} ({\bf x} + {\bf r},t) -\chi ~ \theta ({\bf x},t)~\theta ({\bf x} + {\bf r},t) \rangle  d{\bf r}  \eqno{(9)}  
$$  
where $\chi = \alpha g H/\Delta T$.  

\section{Asymptotically isotropic distributed chaos}

\begin{figure} \vspace{-1cm}\centering
\epsfig{width=.45\textwidth,file=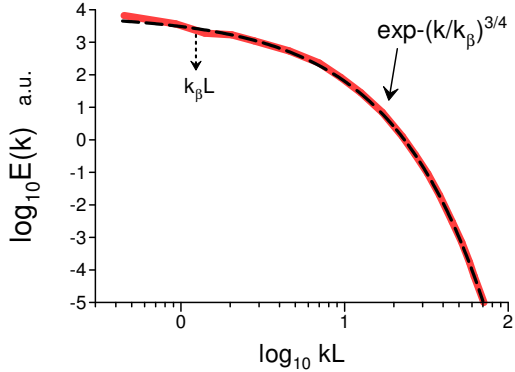} \vspace{-3.87cm}
\caption{Kinetic energy spectrum of the decaying isotropic homogeneous turbulence corresponding to $R_{\lambda} \simeq 80$.} 
\end{figure}
\begin{figure} \vspace{-1cm}\centering
\epsfig{width=.45\textwidth,file=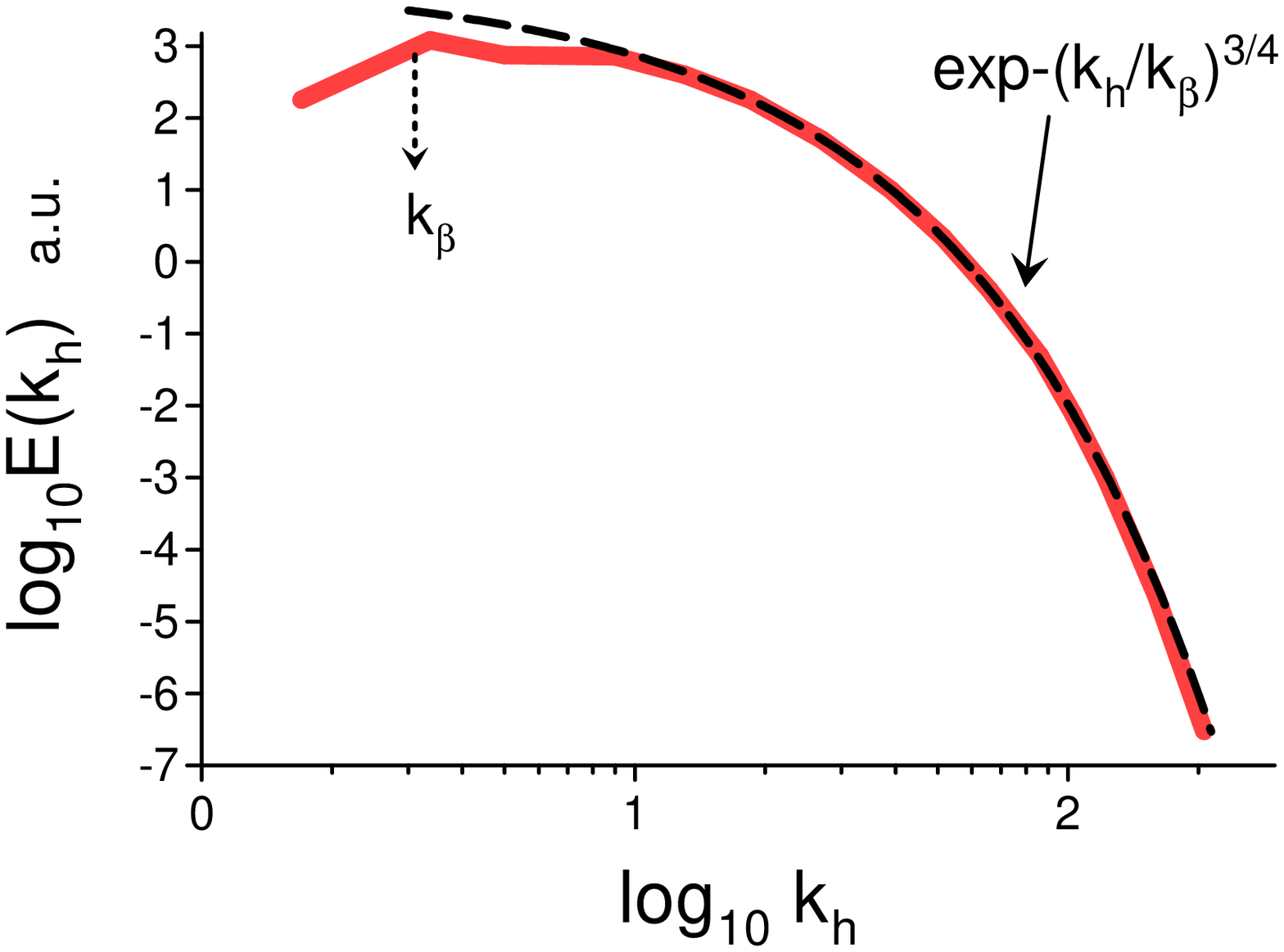} \vspace{-3.55cm}
\caption{The same as in Fig. 1 but for $Ra =10^5$. } 
\end{figure}
\begin{figure} \vspace{-0.65cm}\centering
\epsfig{width=.45\textwidth,file=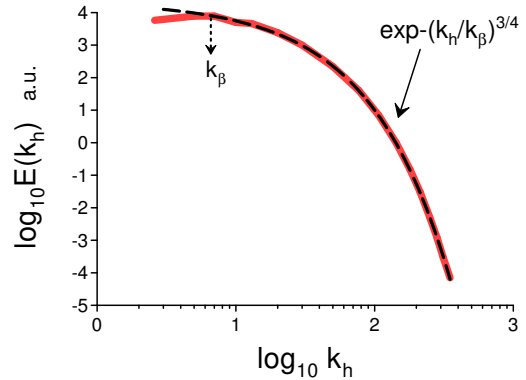} \vspace{-4.22cm}
\caption{The same as in Fig. 1 but for $Ra =10^6$.} 
\end{figure}

   The exponential kinetic energy spectra were also observed at the onset of isotropic homogeneous turbulence \cite{kds}. Therefore, let us begin with the isotropic distributed chaos. \\

   With an increase of the Reynolds (Rayleigh) number the characteristic wavenumber $k_c$ in the exponential spectrum Eq. (4) can become fluctuate. The deterministic chaos will be transformed into distributed chaos with an ensemble-averaging kinetic energy spectrum 
$$
E(k) \propto \int_0^{\infty} P(k_c) \exp -(k/k_c) ~dk_c \eqno{(10)}
$$    
The probability distribution $P(k_c)$ is the main characteristic of the distributed chaos. For the distributed chaos the trajectories are still smooth and, consequently, the kinetic energy spectrum Eq. (10) will be stretched exponential 
$$
E(k) \propto \exp-(k/k_{\beta})^{\beta}  \eqno{(11)}
$$   
where the constant $k_{\beta}$ is a renormalized characteristic wavenumber. Using Eqs. (10) and (11) one can find a universal asymptote of the distribution $P(k_c)$ for large $k_c$ \cite{jon}
$$
P(k_c) \propto k_c^{-1 + \beta/[2(1-\beta)]}~\exp(-ak_c^{\beta/(1-\beta)}) \eqno{(12)}
$$     
here $a$ is a constant.\\

  On the other hand, one can estimate the asymptote of the distribution $P(k_c)$ from physical considerations. Indeed, for the distributed chaos dominated, for instance, by the Birkhoff-Saffman invariant one can use dimensional considerations in order to relate the characteristic velocity fluctuations $u_c$ to the characteristic wavenumber $k_c$
 $$
 u_c^2 \propto I_{BS} ~ k_c^3  \eqno{(13)}
 $$ 
 
 Let us write Eq. (13) in a general form
$$
 u_c^2 \propto k_c^{\gamma}  \eqno{(14)}
 $$ 
 
   If $k_c$ fluctuates, then $u_c$ fluctuates as well. In the case of Gaussian distribution of the characteristic velocity fluctuations $u_c$ \cite{my} 
$$
P(u_c) \propto \exp-\frac{u_c^2}{2\sigma^2},    \eqno{(15)}
$$ 
one obtains from Eq. (14) the asymptotic of $P(k_c)$
$$
P(k_c) \propto k_c^{\frac{\gamma}{2} -1} \exp-bk_c^{\gamma}  \eqno{(16)}
$$
 where $b$ is a constant. 
 
   Comparing Eq. (16) with Eq. (12) we obtain 
 $$
 \beta =\frac{\gamma}{(1+\gamma)}  \eqno{(17)} 
$$   
   
   For the isotropic case, in particular, $\gamma =3$ Eq. (13). Hence the distributed chaos spectrum in this case is 
$$
E(k) \propto \exp-(k/k_{\beta})^{3/4}  \eqno{(18)}
$$

  Assuming certain asymptotic isotropy of the Rayleigh-B\'{e}nard convection (cf Ref. \cite{nath} and references therein) these considerations can be applied to the Rayleigh-B\'{e}nard convection with replacement of the Birkhoff-Saffman integral Eq. (5) by the generalized Birkhoff-Saffman integral Eq. (9), and the estimation Eq. (13) can be replaced by estimation 
$$
 u_c^2 \propto |I_b| ~ k_c^3  \eqno{(19)}
 $$ 
with the same result Eq. (18). The isotropy, in this case, means that the parameter $k_c$ fluctuates similarly in all spatial directions.\\

   For the temperature fluctuations the estimate like Eq. (19) can be written as 
$$
\theta_c^2 \propto \chi^{-1} |I_b| ~ k_c^3  \eqno{(20)}    
$$  
with the same result Eq. (18).

\section{Direct numerical simulations}

 \subsection{Decaying isotropic homogeneous turbulence}
 
  In a recent paper Ref. \cite{xu} results of direct numerical simulations of freely decaying isotropic homogeneous turbulence were reported. The decay was starting at $t=0$ from a turbulent flow in a statistically steady state at the Taylor-Reynolds number $R_{\lambda} \simeq 230$. The initial statistically steady turbulence was generated by a random large-scale energy injection with a constant rate using the Navier–Stokes equations in a spatially periodic box. At an effective time $t=0$ the energy injection was abruptly stopped and a free decay (due to the molecular viscosity) was started. The decay continued till the time $t/T_0 =18$, where $T_0$ is the eddy-turnover time of the turbulence at the initial statistically steady state. At this time (the end of the DNS) the $R_{\lambda} \simeq 80$.\\ 
  
  Figure 3 shows the energy spectrum computed at the end of the DNS (the spectral data were taken from Fig. 6 of the Ref. \cite{xu} and $L(t)$ is the instantaneous integral
scale). The dashed curve indicates correspondence to the spectrum Eq. (18), i.e. the isotropic homogeneous distributed chaos. \\

\begin{figure} \vspace{-1.3cm}\centering
\epsfig{width=.46\textwidth,file=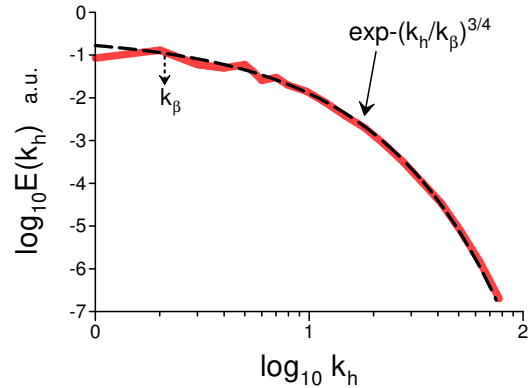} \vspace{-4.2cm}
\caption{Horizontal kinetic energy spectrum spectrum for $Ra =10^6$ but at $z = 0.25$.} 
\end{figure}
\begin{figure} \vspace{-0.5cm}\centering
\epsfig{width=.46\textwidth,file=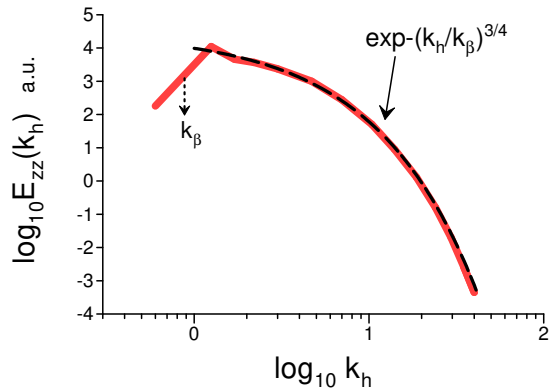} \vspace{-4.1cm}
\caption{ Power spectrum of the {\it vertical} component of the velocity field for $Ra = 3.85\times 10^4$. } 
\end{figure}
\begin{figure} \vspace{-1cm}\centering
\epsfig{width=.46\textwidth,file=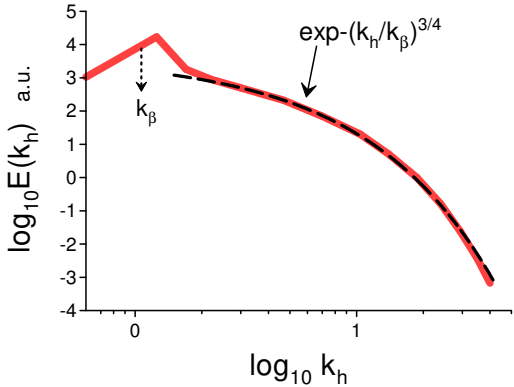} \vspace{-4.1cm}
\caption{Power spectrum of the temperature fluctuations for $Ra = 3.85\times 10^4$. } 
\end{figure}

\subsection{Rayleigh-B\'{e}nard convection at transitional Rayleigh numbers}

  Let us now return to the Rayleigh-B\'{e}nard convection. For the same DNS reported in the Ref. \cite{ytc} (see Section II and Fig. 1) the horizontal kinetic energy spectrum spectra were computed at the midplane ($z = 0.5$)  for $Ra = 10^5$ and $Ra =10^6$. These spectra are shown in Figures 4 and 5 (the spectral data were taken from Fig. 10d of the Ref. \cite{ytc}). The dashed curves indicate correspondence to the spectrum Eq. (18), i.e. the asymptotically isotropic homogeneous distributed chaos (cf Fig. 1 for $Ra= 10^4$).\\

  In paper Ref. \cite{rin} the horizontal kinetic energy spectrum was computed at similar conditions ($Ra =10^6$, $Pr = 1$ and $\Gamma = 5$) as the above-mentioned ones but at the horizontal plane $z=0.25$. This spectrum is shown in Fig. 6 (the spectral data were taken from Fig. 2 of the Ref. \cite{rin}). The dashed curve indicates correspondence to the spectrum Eq. (18), i.e. the isotropic homogeneous distributed chaos. One can see (cf. Fig. 5) that for $Ra =10^6$ the large-scale coherent structures (the spectral peak) determine the value of the $k_{\beta}$ of the asymptotically isotropic distributed chaos.\\
  
   In a DNS reported in a recent paper Ref. \cite{sch} power spectrum of the {\it vertical} component of the velocity field was computed for the Rayleigh-B\'{e}nard convection at $Ra = 3.85\times 10^4$, $Pr = 1$, and $\Gamma =15$. Figure 7 shows this spectrum. The spectral data were taken from Fig. 9 (Dirichlet case) of the Ref. \cite{sch}. The dashed curve indicates correspondence to the spectrum Eq. (18), i.e. the asymptotically isotropic homogeneous distributed chaos. Figure 8 shows the corresponding power spectrum for the temperature fluctuations and the dashed curve indicates correspondence to the spectrum Eq. (18). \\

\begin{figure} \vspace{-0.9cm}\centering
\epsfig{width=.45\textwidth,file=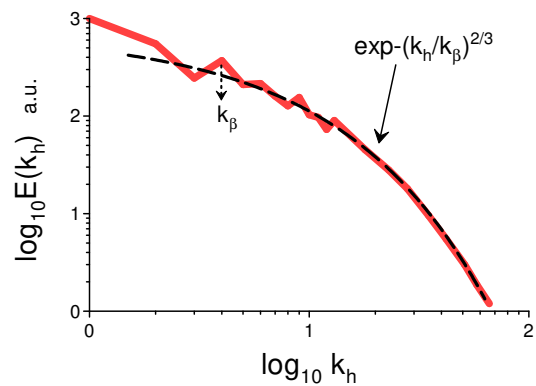} \vspace{-3.98cm}
\caption{Horizontal kinetic energy spectrum obtained in the DNS \cite{par} for $Ra=10^7$ at a statistically stationary state (cf Fig. 2).} 
\end{figure}
\begin{figure} \vspace{-0.5cm}\centering
\epsfig{width=.47\textwidth,file=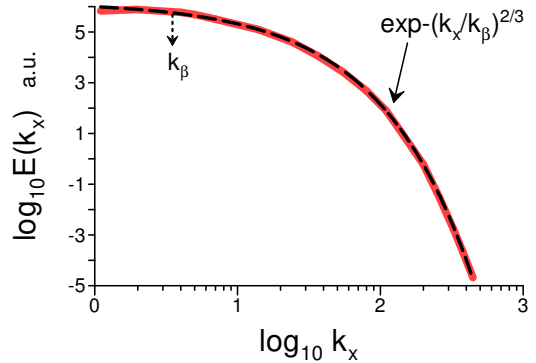} \vspace{-4.3cm}
\caption{ Horizontal kinetic energy spectrum spectrum obtained for $Ra = 1.6\times 10^7$ in DNS reported in Ref. \cite{fon}.} 
\end{figure}
\begin{figure} \vspace{-1cm}\centering
\epsfig{width=.45\textwidth,file=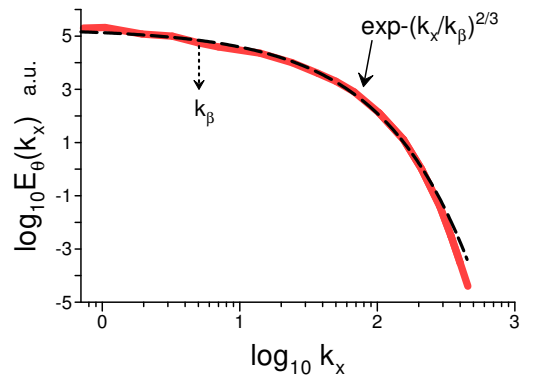} \vspace{-4.3cm}
\caption{As in the Fig. 10 but for the temperature fluctuations} 
\end{figure}

\section{Axisymmetric distributed chaos}

\subsection{Partial asymptotic isotropy}

  With the increase of the Rayleigh number $Ra$ the above-mentioned asymptotic isotropy of the distributed chaos, when the parameter $k_c$ fluctuates similarly in all spatial directions, will be replaced by a partial asymptotic isotropy:
$$
 u_c^2 \propto |I_b| ~ k_{h,c}^2k_{z,c}  \eqno{(21)}
 $$ 
 when the fluctuations of $u_c$ will be caused mainly by the fluctuations of $ k_{h,c}$ and the fluctuations of $k_{z,c}$ in Eq. (21) can be neglected. In this case, $\gamma$ in Eq. (14) should be taken as $\gamma = 2$ and, consequently, $\beta = 2/3$, i.e.
$$
E(k_h) \propto \exp-(k_h/k_{\beta})^{2/3}  \eqno{(22)}
$$    
For the temperature fluctuations, the estimate like Eq. (21) can be written as 
$$
\theta_c^2 \propto \chi^{-1} |I_b| ~ k_{h,c}^2k_{z,c} \eqno{(23)}    
$$  
with the same result Eq. (22).

\subsection{Direct numerical simulation}

\begin{figure} \vspace{-0.8cm}\centering
\epsfig{width=.45\textwidth,file=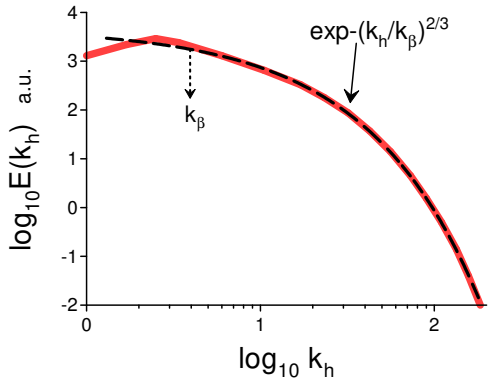} \vspace{-4.14cm}
\caption{Horizontal kinetic energy spectrum obtained in the DNS \cite{stev} for $Ra=2 \times 10^7$ at the midplane of the DNS domain.} 
\end{figure}
\begin{figure} \vspace{-1cm}\centering
\epsfig{width=.45\textwidth,file=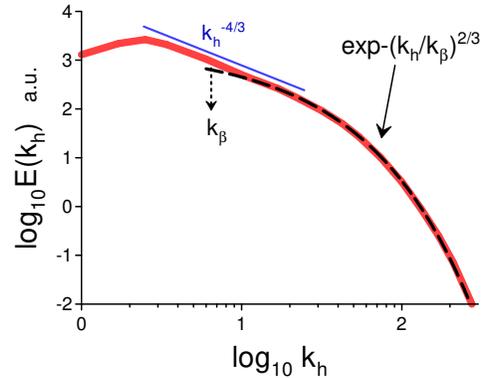} \vspace{-3.95cm}
\caption{As in Fig. 12 but for $Ra = 10^8$.} 
\end{figure}
\begin{figure} \vspace{-0.5cm}\centering
\epsfig{width=.45\textwidth,file=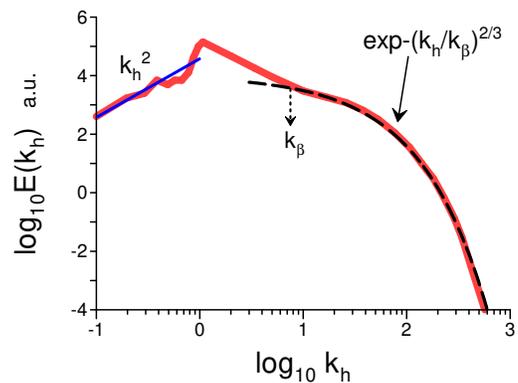} \vspace{-3.95cm}
\caption{As in Fig. 13 but for $z=0.062$.} 
\end{figure}

  In Section II results of direct numerical simulations \cite{par} for a moderately large Rayleigh number $Ra = 10^7$
at the onset of the DNS (at the $t=13$ in the terms of the DNS) indicated the deterministic chaos with the exponential spectrum Fig. 2. Now let us consider the {\it statistically stationary state} for this DNS at $t \sim 160$ in the terms of the DNS. Figure 9 shows the corresponding horizontal kinetic energy spectrum spectrum (the spectral data were taken from Fig. 3a of the Ref. \cite{par}). The dashed curve indicates the stretched exponential spectrum Eq. (22) corresponding to the axisymmetric (partially isotropic) distributed chaos. 

It is interesting that $k_{\beta}$ corresponds to a large-scale spectral peak, in this case, i.e. the distributed chaos is dominated by some large-scale coherent structures. And indeed, the authors of the Ref. \cite{par} noticed that the convective (descending) cold and (rising) hot plumes organize themselves into certain large-scale clusters. Most of the kinetic energy of the fluid motion is concentrated in these plumes. \\

\begin{figure} \vspace{-1cm}\centering
\epsfig{width=.44\textwidth,file=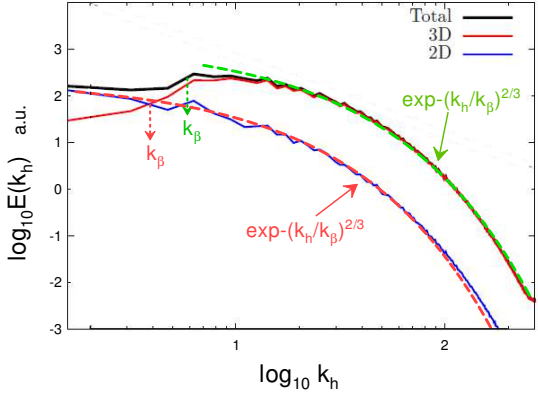} \vspace{-3.7cm}
\caption{Horizontal kinetic energy spectra for $Ra=3\times 10^7$, and $Ta = 10^8$.} 
\end{figure}

   Figure 10 shows the horizontal kinetic energy spectrum obtained for $Ra = 1.6\times 10^7$, $Pr=1$, and $\Gamma = 2\pi$ in DNS reported in a recent Ref. \cite{fon}. The spectral data were taken from Fig. 15b (the midplane of the domain) of the Ref. \cite{fon}. As in the DNS reported in the Ref \cite{par} the no-slip boundary conditions were taken for the velocity field on the top and bottom boundaries. The dashed curve indicates the stretched exponential spectrum Eq. (22) corresponding to the axisymmetric (partially isotropic) distributed chaos. 
  
  Figure 11 shows the power spectrum of the temperature fluctuations obtained in the same DNS. The dashed curve indicates the stretched exponential spectrum Eq. (22).\\ 
  
  In a recent paper Ref. \cite{stev} results of DNS for the Rayleigh numbers $Ra = 2\times 10^7$ and $Ra = 10^8$ were reported. The DNS were performed for a very high $\Gamma = 64$ ($Pr =1$). The no-slip boundary condition on the top and bottom boundaries and periodic boundary conditions in all horizontal directions were taken for the velocity field. \\
  
  Figures 12 and 13 show the horizontal kinetic energy spectra obtained for $Ra=2 \times 10^7$ and $Ra = 10^8$ for the midplane ($z = 0.5$) of the DNS domain. The spectral data were taken from Figs. 2f and 2b of the Ref. \cite{stev} respectively. The dashed curves indicate the stretched exponential spectrum Eq. (22) corresponding to the axisymmetric (partially isotropic) distributed chaos. 
  
   It should be noted that for $Ra = 10^8$ an indication of a scaling appeared at the large scales because the trajectories cannot be sufficiently smoothed by the dissipative effects at these scales (see Introduction). The value of the large-scale scaling exponent `-4/3' can be related to the superstructures observed in the DNS (see, for instance, Ref. \cite{ber} and references therein).
   
   Figure 14 shows the horizontal kinetic energy spectrum also obtained for $Ra = 10^8$ but at $z =0.062$ (the spectral data were taken from Fig. 8 of Ref. \cite{fran}). The dashed curve indicates the stretched exponential spectrum Eq. (22) corresponding to the axisymmetric (partially isotropic) distributed chaos. The straight line indicates appearance of the Saffman-like spectrum \cite{saf},\cite{dav} $E(k_h) \propto |I_b| k_h^2$ for the small wavenumbers, where the Birkhoff-Saffman invariant $I_{BS}$ has been replaced by the generalized one $I_b$.\\
  
  The addition of a rotation with a constant angular velocity $\bm{\Omega}=\Omega\bm{e}_z$, i.e. addition of the Coriolis force into the equation Eq. (6) (in a non-dimensional form the term: $\sqrt{Ta}\,\bf{u}\times \bf{e}_z$, where $Ta$ is the Taylor number $Ta=4\Omega^2H^4/\nu^2$), is compatible with the previous consideration (cf Ref. \cite{dav}).\\ 
 
\begin{figure} \vspace{-0.95cm}\centering
\epsfig{width=.45\textwidth,file=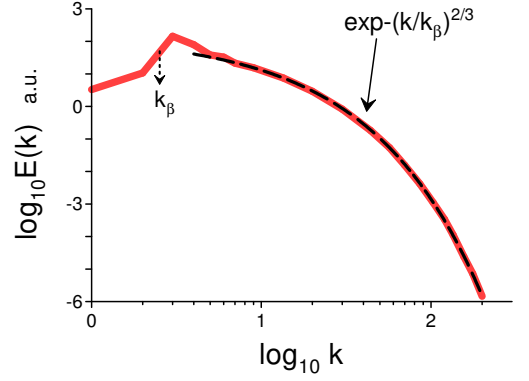} \vspace{-4cm}
\caption{Horizontal kinetic energy  spectrum of the stably stratified flow for $Ri =0.01$ and $Pr = 1$.} 
\end{figure}
\begin{figure} \vspace{-0.45cm}\centering
\epsfig{width=.45\textwidth,file=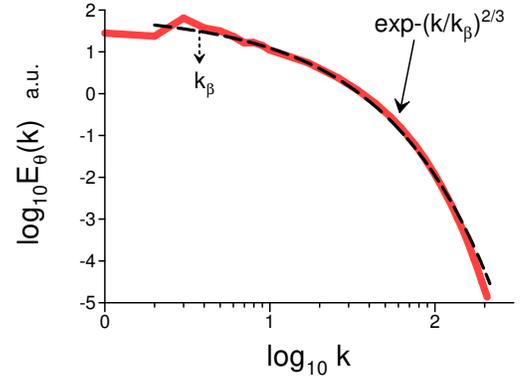} \vspace{-4.1cm}
\caption{As in Fig. 16 but for the temperature fluctuations.} 
\end{figure}

  In a recent paper Ref. \cite{fav} results of direct numerical simulations of the fast-rotating Rayleigh-B\'{e}nard convection were reported. The authors of the Ref. \cite{fav} defined the z-averaged horizontal (2D) flow as
$$
\left<u_x\right>_z(x,y) = \int u_x(x,y,z)  dz,   
$$
$$
\left<u_y\right>_z (x,y) = \int u_z(x,y,z) dz 
$$
and 3D (z-depended) fluctuations as
$$
u_x'(x,y,z)  = u_x(x,y,z)-\left<u_x\right>_z(x,y),
$$
$$
u_y'(x,y,z)  = u_y(x,y,z)-\left<u_y\right>_z(x,y)
$$
Then the horizontal 2D kinetic energy was defined as
$$
K_{2D}=\frac{1}{2\Gamma^2}\iint \left(\left<u_x\right>_z^2+\left<u_y\right>_z^2\right)  dx  dy
$$
and the 3D kinetic energy associated with the fluctuations as
$$
K_{3D}=\frac{1}{2\Gamma^2}\iiint \left(u_x'^2+u_y'^2+u_z^2\right)  dx  dy  dz  
$$

The total kinetic energy is $K_{tot} = K_{3D}+ K_{2D}$. \\

   Figure 15 shows the averaged over time ($0.3 < t < 0.4$, in the terms of the Ref. \cite{fav}) and depth ($0 < z < 1$) horizontal kinetic energy spectra vs $k_h$ for $Ra=3\times 10^7$, $Ta = 10^8$, $Pr =1$, and $\Gamma =4$. The spectral data were taken from Fig. 3a of the Ref. \cite{fav}. The dashed curves indicate the spectrum Eq. (22) (axisymmetric distributed chaos).

\section{Discussion}

\subsection{Stably stratified flows}

   The above-discussed approach can be also applied to stably stratified flows. The sign before the term $\chi ~ \theta ({\bf x},t)~\theta ({\bf x} + {\bf r},t) \rangle$  in the generalized Birkhoff-Saffman invariant Eq. (9) should be changed for this case (cf Ref. \cite{kcv}). \\
  
  In the Ref. \cite{kcv} results of direct numerical simulations of the stably stratified flows generated by an external large-scale random forcing were reported. The DNS were performed in a cubic box with periodic boundary conditions in all (three) directions for $Pr =1$ and the Richardson number $Ri = \alpha g \Delta T H/ u_{\textrm{rms}}^2= 0.01$.\\ 
  
    Figure 16 shows the horizontal kinetic energy spectrum computed for this simulation (the spectral data were taken from Fig. 2a of the Ref. \cite{kcv}). The dashed curve indicates the stretched exponential spectrum Eq. (22) (axisymmetric distributed chaos).  
    
    Figure 17 shows the corresponding power spectrum of the temperature fluctuations. The dashed curve indicates the stretched exponential spectrum Eq. (22) (axisymmetric distributed chaos).  \\
 
 \subsection{Upper troposphere and lower stratosphere}
 
\begin{figure} \vspace{-1.5cm}\centering
\epsfig{width=.45\textwidth,file=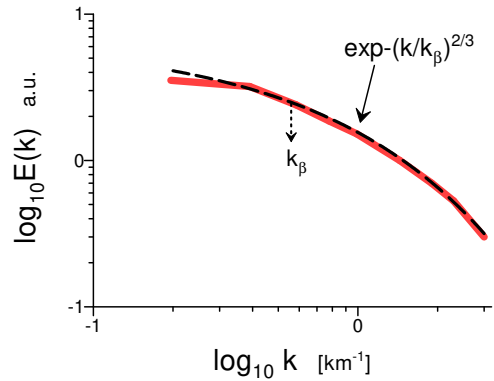} \vspace{-3.8cm}
\caption{Power spectrum of the velocity component longitudinal relative to the mean wind. The data were averaged over the turbulent episodes both for the upper troposphere and lower stratosphere.} 
\end{figure}
\begin{figure} \vspace{-0.5cm}\centering
\epsfig{width=.45\textwidth,file=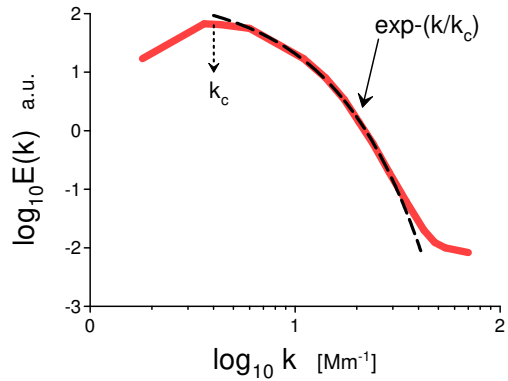} \vspace{-3.8cm}
\caption{Power spectrum of the Doppler velocity for the quiet-Sun region located at the disk center} 
\end{figure}
\begin{figure} \vspace{-0.5cm}\centering
\epsfig{width=.45\textwidth,file=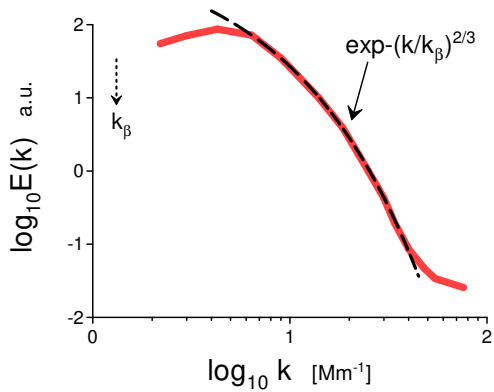} \vspace{-3.9cm}
\caption{Power spectrum of the Doppler velocity for the active solar region AR11768} 
\end{figure}
 
  The Global Atmospheric Sampling Program - GASP (flight-level aircraft measurements), made a considerable contribution to the understanding of atmospheric turbulence. A special accent was made on the mesoscale processes (see, for instance, the seminal paper \cite{ng}). The so-called small-scale (in this sense) turbulent episodes, when the GASP airplane encountered moderate or strong atmospheric turbulence, attracted much less attention (maybe because the observed in these episodes spectra did not fit the scaling power laws). \\
  
    Figure 18 shows the power spectrum of the velocity component longitudinal relative to the mean wind. The data were averaged over the moderate turbulent episodes both for the upper troposphere and lower stratosphere, where the power spectra show no significant difference (the spectral data were taken from Fig. 2 of the Ref. \cite{ng}). The dashed curve indicates the stretched exponential spectrum Eq. (22).

 \subsection{In the solar photosphere}
    
    The distributed chaos was already recognized in the solar photosphere for the magnetic field fluctuations in an active region \cite{ber1}. Of course, the theoretical consideration of the solar photosphere should include radiative transfer, compressible magnetohydrodynamics, and so on. However, one can expect that all these effects can be compatible with the above consideration. Therefore, one can try to apply this consideration to the observational data.\\
    
    In a recent paper Ref. \cite{cha} results of the high-resolution spectropolarimetric observations obtained with the balloon-borne Sunrise observatory (the largest balloon-borne solar telescope located above 99\% of the atmosphere) were reported. The observations were performed with the Imaging Magnetograph eXperiment (IMaX) and with the Sunrise Filter Imager (SuFI) in 2009 June and 2013 June. On 2009 June 9 the telescope was pointed to a quiet-Sun region located at disk center, whereas on 2013 June 12 the telescope was pointed to the trailing part of the active solar region AR11768. Using the observational data the authors of the Ref. \cite{cha} computed the Doppler velocity power spectra. 
  
   Figure 19 shows the power spectrum of the Doppler velocity for the quiet-Sun region located at the disk center (the spectral data were taken from Fig. 3 of the Ref. \cite{cha}). The dashed curve indicates the exponential spectrum Eq. (4) (deterministic chaos). 
   
   Figure 20 shows the power spectrum of the Doppler velocity for the active solar region AR11768 (the spectral data were taken from Fig. 3 of the Ref. \cite{cha}). The dashed curve indicates the stretched exponential spectrum Eq. (22).


\begin{thebibliography}{99}
\bibitem{hcl}  F. Heslot, B. Castaing, A. Libchaber, Phys. Rev. A, {\bf 36}, 5870 (1987)
\bibitem{lorenz} E.N. Lorenz, J. Atmos. Sci., {\bf 20}, 130 (1963)
\bibitem{fm} U. Frisch and R. Morf, Phys. Rev., {\bf 23}, 2673 (1981)
\bibitem{oh} N. Ohtomo, K. Tokiwano, Y. Tanaka et. al., J. Phys. Soc.
Jpn., {\bf 64}, 1104 (1995)
\bibitem{mm1} J. E. Maggs and G. J. Morales, Phys. Rev. Lett., {\bf 107},185003 (2011) 
\bibitem{mm2} J. E. Maggs and G. J. Morales, Phys. Rev. E {\bf 86}, 015401(R) (2012)
\bibitem{kds} S. Khurshid, D.A. Donzis and K.R. Sreenivasan, Phys. Rev. Fluids, {\bf 3}, 082601(R) (2018)
\bibitem{ytc} M. Yan, S.M. Tobias, and M.A. Calkins, J. Fluid Mech., {\bf 915}, A15 (2021)
\bibitem{par} A. Parodi, J. von Hardenberg, G. Passoni, A. Provenzale, and E.A. Spiegel, Phys. Rev. Lett., {\bf 92}, 194503 (2004)
\bibitem{my} A. S. Monin, A. M. Yaglom, Statistical Fluid Mechanics, Vol. II: Mechanics of Turbulence (Dover Pub. NY, 2007)
\bibitem{bir} G. Birkhoff, Commun. Pure Appl. Math., {\bf 7}, 19 (1954)
\bibitem{saf} P. G. Saffman, J. Fluid. Mech., {\bf 27}, 551 (1967)
\bibitem{dav} P.A. Davidson, J. Fluid Mech., {\bf 663}, 268 (2010)
\bibitem{kcv} A. Kumar, A.G. Chatterjee and M.K. Verma, Phys. Rev. E, {\bf 90}, 023016 (2014)
\bibitem{jon} D.C. Johnston, Phys. Rev. B, {\bf 74}, 184430 (2006)
\bibitem{nath} D. Nath, A. Pandey, A. Kumar, and M.K. Verma, Phys. Rev. Fluids, {\bf 1}, 064302 (2016)
\bibitem{xu} P-F. Yang, A. Pumir, and H. Xu, New J. Phys., {\bf 20} 103035 (2018)
\bibitem{rin} F. Rincon, J. Fluid. Mech.,  {\bf 563}, 43 (2006)
\bibitem{sch} P.P. Vieweg, J.D. Scheel, J. Schumacher, Phys. Rev. Research, {\bf 3}, 013231 (2021)
\bibitem{stev} R.J.A.M. Stevens, A. Blass, X.. Zhu, R Verzicco, and D Lohse, Phys. Rev. Fluids, {\bf 3}, 041501(R) (2018)
\bibitem{ber} A. Bershadskii, Physica A, {\bf 206}, 120 (1994)
\bibitem{fran} A. Franken, E-print, University of Twente, (2017) https://essay.utwente.nl/73950/
\bibitem{fon} M. Fontana, O.P. Bruno, P.D. Mininni, and P. Dmitruk, Comp. Phys. Comm., {\bf 256}, 107482 (2020)
\bibitem{fav} B. Favier, C. Guervilly, and E. Knobloch, J.  Fluid Mech., {\bf 864}, R1 (2019)
\bibitem{ng} G.D. Nastrom and K.S. Gage, Tellus A, {\bf 35} 383 (1983)
\bibitem{ber1} A. Bershadskii, Res. Notes AAS, {\bf 4}, 10 (2020)
 \bibitem{cha} L. Yelles Chaouche, R.H. Cameron, S.K. Solanki, et al.,  A\&A, {\bf 644}, A44 (2020)






\end{thebibliography}
\end{document}